\documentclass{andromedaone}       

\journal{BSM}%Letters in High Energy Physics}
\vol{2021}
\jyear{Egypt}
\pages{Zewail} 

\received{xx January 2018}
\published{xx March 2018}

\usepackage{amsmath,amssymb}

\newcommand{\rR}{\rho_R}
\newcommand{\rbh}{\rho_\text{BH}}
\newcommand{\gs}{g_\star}
\newcommand{\gss}{g_{\star s}}
\newcommand{\Tbh}{T_\text{BH}}
\newcommand{\Tbhin}{T_\text{BH}^\text{in}}
\newcommand{\Mbh}{M_\text{BH}}
\newcommand{\tev}{t_\text{ev}}
\newcommand{\Tev}{T_\text{ev}}
\newcommand{\Tbev}{\bar T_\text{ev}}
\newcommand{\Min}{M_\text{in}}
\newcommand{\Tin}{T_\text{in}}
\newcommand{\tin}{t_\text{in}}
\newcommand{\ndm}{n_\text{DM}}
\newcommand{\Ndm}{N_\text{DM}}
\newcommand{\gdm}{g_\text{DM}}
\newcommand{\Tpev}{{T^\prime_\text{ev}}}
\newcommand{\Cr}{\mathcal{C}_\rho}
\newcommand{\Cn}{\mathcal{C}_n}

\newcommand{\nin}{n_\text{in}}

\newcommand{\mdm}{m_\text{DM}}
\newcommand{\Trh}{T_\text{rh}}
\newcommand{\Tmax}{T_\text{max}}

\begin{document}

\title{Gravitational Dark Matter and Primordial Black Holes}

\author{Nicolás Bernal}
\address{Centro de Investigaciones, Universidad Antonio Nariño, Carrera 3 Este \# 47A-15, Bogotá, Colombia}

\begin{abstract}
Let's be frank: we all love WIMP dark matter (DM).
However, thermal equilibrium may not be reached between the visible and dark sectors.
The only interaction guaranteed between them is gravity.
Gravitational DM could be generated by the Hawking evaporation of primordial black holes.
It is typically assumed that after production, DM abundance freezes and remains constant.
However, thermalization and number-changing processes in the dark sector can have a strong
impact, in particular enhancing the DM population by several orders of magnitude.
Here we review the impact of self-interactions from general arguments such as the conservation of energy and entropy, independently from the underlying particle physics details of the dark sector.
Additionally, we study the interplay with other purely gravitational DM production mechanisms, such as the scattering of standard model states or inflatons, mediated by the $s$-channel exchange of gravitons.
We show that the latter mechanism sets strong bounds, excluding large regions of the parameter space favored by primordial black hole production.
\end{abstract}

\maketitle


%%%%%%%%%%%%%%%%%%%%%%%%%%%%%%%%%%%%%%%%%%%%%%%%%%%%%%%%%%
\section{Introduction}
%%%%%%%%%%%%%%%%%%%%%%%%%%%%%%%%%%%%%%%%%%%%%%%%%%%%%%%%%%
The nature of dark matter (DM) is currently one of the most intriguing questions of fundamental physics.
Its existence has been proven beyond major doubts from several astrophysical and cosmological evidences.
In the last decades, weakly interacting massive particles (WIMPs), with masses and couplings at the electroweak scale, have been the leading DM production paradigm~\cite{Arcadi:2017kky}.
However, the increasingly strong observational constraints on DM (e.g., null results from direct and indirect detection experiments, together with production at colliders) are urging the quest for alternative scenarios.

Other possibilities exist.
For example, the thermal history of the universe could deviate from the standard radiation-dominated picture~\cite{Allahverdi:2020bys}.
Alternatively, DM could couple very feebly with the standard model (SM), so that the two sectors never reach chemical equilibrium, as in the case of the feebly interacting massive particle (FIMP) paradigm~\cite{McDonald:2001vt, Choi:2005vq, Hall:2009bx, Elahi:2014fsa, Bernal:2017kxu}.
It is interesting to note that even if the strength of the couplings between the visible and dark sectors is an open question, the gravitational interaction is the {\it irreducible} force between them.
Additionally, it has been pointed out that the whole observed DM abundance can be successfully produced by purely gravitational interactions~\cite{Garny:2015sjg, Tang:2017hvq, Garny:2017kha, Bernal:2018qlk, Ema:2016hlw, Ema:2018ucl, Mambrini:2021zpp, Bernal:2021kaj, Barman:2021ugy}.

On the one hand, one possibility widely studied in the literature corresponds to the DM radiation via Hawking evaporation~\cite{Hawking:1974sw} of primordial black holes (PBH)~\cite{Carr:1974nx, Carr:2009jm, Carr:2020gox}.
It is well known that during its evaporation, PBHs radiate {\it all} particles lighter than its mass, and therefore visible but also hidden sector states~\cite{Green:1999yh, Khlopov:2004tn, Dai:2009hx, Fujita:2014hha, Allahverdi:2017sks, Lennon:2017tqq, Morrison:2018xla, Hooper:2019gtx, Chaudhuri:2020wjo, Masina:2020xhk, Baldes:2020nuv, Gondolo:2020uqv, Bernal:2020kse, Bernal:2020ili, Bernal:2020bjf}.
A particularly interesting range for the initial PBH mass corresponds to $\sim 0.1$~g to $\sim 100$ tones, because in this case they completely evaporate before the onset of Big Bang nucleosynthesis (BBN), and are therefore poorly constrained (see however Refs.~\cite{Papanikolaou:2020qtd, Domenech:2020ssp}).
Finally, we note that an incomplete evaporation of PBHs renders them viable DM candidates~\cite{Carr:2016drx, Carr:2020xqk, Green:2020jor, Villanueva-Domingo:2021spv}.
However, even if gravity is the only interaction mediating between the visible and dark sectors, DM could feature sizable (i.e., non-gravitational interactions).
These self-interactions could bring DM into both kinetic and chemical equilibrium, having a strong impact on the final DM relic abundance.

On the other hand, other purely gravitational DM production mechanism exists.
For instance, SM states could produce DM particles via the 2-to-2 scattering mediated by the $s$-channel exchange of gravitons~\cite{Garny:2015sjg, Tang:2017hvq, Garny:2017kha, Bernal:2018qlk}.
Additionally, during the heating era, DM could also be created by inflaton scatterings, again via the mediation of gravitons~\cite{Ema:2016hlw, Ema:2018ucl, Mambrini:2021zpp, Bernal:2021kaj, Barman:2021ugy}.

The paper is organized as follows.
In section~\ref{sec:PBH} we briefly review key aspects of PBH creation and evaporation, whereas in section~\ref{sec:DM} the DM production from PBH evaporation is discussed.
Section~\ref{sec:SI} is devoted to the impact of DM self-interactions on its relic abundance.
The constraints coming from other purely gravitational DM production mechanisms are presented in Section~\ref{sec:gravity}.
Finally, the conclusions are presented in Section~\ref{sec:conclusions}.

%%%%%%%%%%%%%%%%%%%%%%%%%%%%%%%%%%%%%%%%%%%%%%%%%%%%%%%%%%
\section{Primordial Black Holes: Formation and Evaporation} \label{sec:PBH}
%%%%%%%%%%%%%%%%%%%%%%%%%%%%%%%%%%%%%%%%%%%%%%%%%%%%%%%%%%
PBHs produced during the radiation dominated epoch, when the SM plasma has a temperature $T=\Tin$, have an initial mass $\Min$ of the order of the enclosed mass in the particle horizon~\cite{Carr:2009jm, Carr:2020gox, Masina:2020xhk, Gondolo:2020uqv}
\begin{equation}
    \Min \equiv \Mbh(\Tin) = \frac{4\pi}{3}\, \gamma\, \frac{\rR(\Tin)}{H^3(\Tin)}\,,
\end{equation}
where $\gamma\simeq 0.2$, $\rR(T)\equiv\frac{\pi^2}{30}\,\gs\,T^4$ is the SM radiation energy density with $\gs(T)$ the number of relativistic degrees of freedom contributing to $\rR$~\cite{Drees:2015exa}, and $H^2(T)=\frac{\rho(T)}{3M_P^2}$ is the squared Hubble expansion rate in terms of the total energy density $\rho(T)$, with $M_P$ the reduced Planck mass.

Once a PBH is formed, the evaporation process starts to radiate particles lighter than the BH temperature $\Tbh = M_P^2/\Mbh$~\cite{Hawking:1974sw}, leading to a mass loss rate~\cite{Page:1976df, Gondolo:2020uqv}
\begin{equation}\label{eq:dMdt}
    \frac{d\Mbh}{dt} = -\frac{\pi\, \gs(\Tbh)}{480}\frac{M_P^4}{\Mbh^2}\,.
\end{equation}
If the universe remains dominated by SM radiation along the BH lifetime, it follows that SM plasma temperature when the BH has completely faded away is
\begin{equation}\label{eq:Tev}
    \Tev \equiv T(\tev)\simeq \left(\frac{9\,\gs(\Tbh)}{10240}\right)^\frac14 \left(\frac{M_P^5}{\Min^3}\right)^\frac12.
\end{equation}
However, if the PBH component dominates at some point the total energy density of the universe, the SM temperature just after the complete evaporation of PBHs is $\Tbev=\frac{2}{\sqrt{3}}\Tev$.  

The total number $N_j$ of the species $j$ of mass $m_j$ emitted during the PBH evaporation is given by
\begin{equation}\label{eq:N}
    N_j=\frac{15\,\zeta(3)}{\pi^4}\frac{g_j\,\Cn}{\gs(\Tbh)} \times
    \begin{cases}
        \left(\frac{\Min}{M_P}\right)^2\qquad\text{for}\quad m_j\leq\Tbhin\,,\\[8pt]
        \left(\frac{M_P}{m_j}\right)^2 \qquad\text{for}\quad m_j\geq\Tbhin\,,
    \end{cases}
\end{equation}
where $\Tbhin\equiv\Tbh(t=\tin)$ is the initial PBH temperature, and $\Cn = 1$ or $3/4$ for bosonic or fermionic species, respectively.
We note that the radiated particles are relativistic, with a mean energy $\langle E_j\rangle = 6\times \max(m_j,\,\Tbhin)$.

The initial PBHs abundance is characterized by the dimensionless parameter $\beta$ 
\begin{equation}
    \beta\equiv\frac{\rbh(\Tin)}{\rR(\Tin)}\,,
\end{equation}
which corresponds to the ratio of the initial PBH energy density to the SM energy density at the time of formation $T=\Tin$.
We note that $\beta$ steadily grows until evaporation, given the fact that BHs scale like non-relativistic matter ($\rbh \propto a^{-3}$, with $a$ being the scale factor), while $\rR \propto a^{-4}$.
A matter-dominated era (i.e., a PBH domination) can be avoided if $\rbh\ll\rR$ at all times, or equivalently if $\beta \ll \Tev/\Tin$\,.

It has been recently pointed out that the production of gravitational waves (GW) induced by large-scale density perturbations underlain by PBHs could lead to a backreaction problem.
However, it could be avoided if the energy contained in GWs never overtakes the one of the background universe~\cite{Papanikolaou:2020qtd}:
\begin{equation} \label{eq:GW}
    \beta < 10^{-4}\left(\frac{10^9~\text{g}}{\Min}\right)^{1/4}.
\end{equation}
Additionally, stronger constrains on the amount of produced GWs come from the requirement that the universe must be radiation dominated at the BBN epoch~\cite{Domenech:2020ssp}, and therefore
\begin{equation} \label{eq:GW2}
    \beta < 3.3\times 10^{-8} \left(\frac{\gamma}{0.2}\right)^{-1/2} \left(\frac{\Min}{10^4~\text{g}}\right)^{-7/8}.
\end{equation}

Finally, PBH evaporation produces all particles, and in particular extra radiation that can modify successful BBN predictions.
To avoid it, we require PBHs to fully evaporate before BBN time, i.e., $\Tev>T_\text{BBN}\simeq 4$~MeV~\cite{Sarkar:1995dd, Kawasaki:2000en, Hannestad:2004px, DeBernardis:2008zz, deSalas:2015glj}, which translates into an upper bound on the initial PBH mass.
On the opposite side, a lower bound on $\Min$ can be set once the upper bound on the inflationary scale is taken into account: $H_{I} \leq 2.5\times10^{-5}M_P$~\cite{Akrami:2018odb}.
Therefore, one has that
\begin{equation}\label{eq:PBHmass}
    0.1~\text{g}\lesssim \Min \lesssim 2\times 10^8~\text{g}\,.
\end{equation}

It is interesting to note that the upper bound on the inflationary scale was used assuming an instantaneous decay of the inflaton, and therefore that the reheating temperature $\Trh$ corresponds the highest temperature reached by the SM thermal bath.
As the inflaton does not decay instantaneously, the SM bath temperature may rise to a value $\Tmax$ much higher than $\Trh$~\cite{Giudice:1999fb, Giudice:2000ex}.
The value of $\Tmax$ depends both on the inflaton decay width and on the inflationary scale~\cite{Barman:2021ugy}.
Now, using again the upper bound on the inflationary scale together with the requirement of fully evaporation before BBN implies that
\begin{equation}
    \frac{\Tmax}{\Trh} \lesssim 100\,.
\end{equation}
Higher ratios $\Tmax/\Trh$ are naturally possible, but incompatible with PBHs that fully evaporate before the onset of BBN.

%%%%%%%%%%%%%%%%%%%%%%%%%%%%%%%%%%%%%%%%%%%%%%%%%%%%%%%%%%
\section{Dark Matter Production from Primordial Black Holes} \label{sec:DM}
%%%%%%%%%%%%%%%%%%%%%%%%%%%%%%%%%%%%%%%%%%%%%%%%%%%%%%%%%%

The whole observed DM relic abundance could have been Hawking radiated by PBHs.
The DM production can be analytically computed in two limiting regimes where PBHs dominate or not the energy density of the universe, and will be presented in the following.

Firstly, the DM yield $Y_\text{DM}$ is defined as the ratio of the DM number density $\ndm$ and the SM entropy density $s(T) \equiv \frac{2\pi^2}{45} \gss(T)\, T^3$, where $\gss(T)$ is the number of relativistic degrees of freedom contributing to the SM entropy~\cite{Drees:2015exa}.
In a radiation dominated universe, the DM yield produced by Hawking evaporation of PBHs can be estimated by
\begin{equation}\label{eq:YdmRD}
    Y_\text{DM} \equiv \frac{\ndm(T_0)}{s(T_0)}
    = N_\text{DM}\,\frac{\nin}{s(\Tin)}
    = \frac34\,\Ndm\,\beta\,\frac{\gs(\Tbh)}{\gss(\Tbh)}\,\frac{\Tin}{\Min}\,,
\end{equation}
with $T_0$ the SM temperature at present, and where the conservation of SM entropy was used.
Additionally, $\Ndm$ is the total number of DM particles emitted by a PBH and is given by Eq.~\eqref{eq:N}.

Alternatively, PBHs can dominate the universe energy density before their decay.
In that case, the DM yield is instead
\begin{equation}\label{eq:YMD}
    Y_\text{DM}\equiv\frac{\ndm(T_0)}{s(T_0)}
    =\frac{\ndm(\Tbev)}{s(\Tbev)}
    \simeq \frac34\, \Ndm\, \frac{\gs(\Tbh)}{\gss(\Tbh)}\, \frac{\Tbev}{\Min}\,,
\end{equation}
using again the conservation of the SM entropy {\it after} the PBHs have completely evaporated, and assuming an {\it instantaneous} evaporation of the PBHs at $t=\tev\simeq\tau$.
We notice that, as expected, the DM yields in the radiation dominated (Eq.~\eqref{eq:YdmRD}) and matter dominated (Eq.~\eqref{eq:YMD}) eras coincide in the limit $\beta \to \Tev/\Tin$, with $\Tbev = \Tev$.

%%%%%%%%%%%%%%%%%%%%%%%%%%%%%%%%%%%%%%%%%%%%%%%%%%%%%%
\begin{figure}
	\centering
	\includegraphics[scale=0.8]{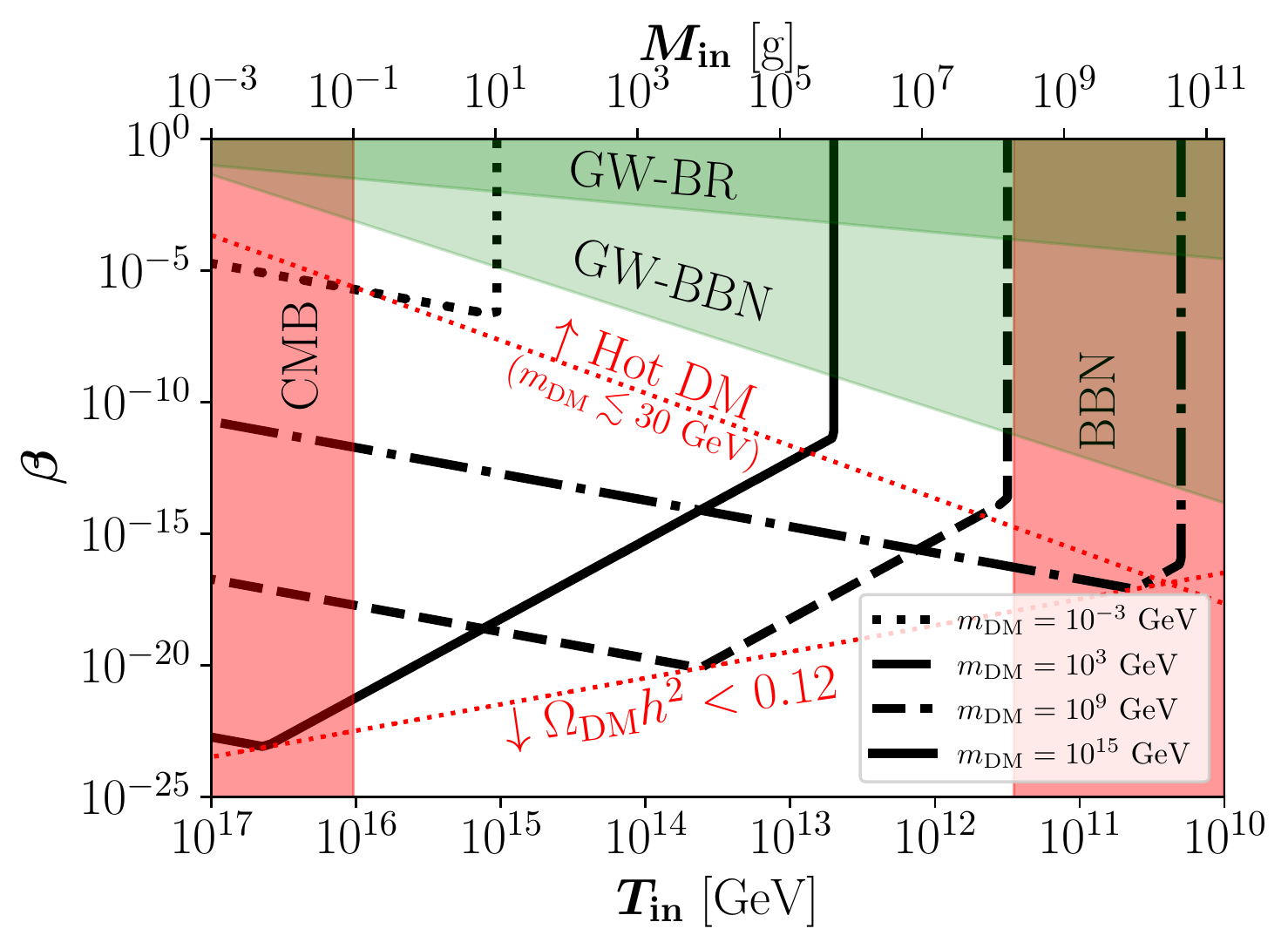}
	\caption{Parameter space reproducing the observed DM abundance (thick black lines) from PBH evaporation, {\it without} DM self-interactions.
	The shaded areas are excluded by different observables described in the text.
	The hot DM bound {\it only} applies to $\mdm \lesssim 30$~GeV.}
	\label{fig:Ti-beta}
\end{figure} 
%%%%%%%%%%%%%%%%%%%%%%%%%%%%%%%%%%%%%%%%%%%%%%%%%%%%%%
To reproduce the observed DM relic abundance $\Omega_\text{DM} h^2\simeq 0.12$~\cite{Aghanim:2018eyx}, the DM yield has to be fixed so that $\mdm\,Y_\text{DM} = \Omega_\text{DM} h^2 \, \frac{1}{s_0}\,\frac{\rho_c}{h^2} \simeq 4.3 \times 10^{-10}$~GeV, where $\rho_c \simeq 1.1 \times 10^{-5} \, h^2$~GeV/cm$^3$ is the critical energy density, and $s_0\simeq 2.9\times 10^3$~cm$^{-3}$ is the entropy density at present~\cite{Aghanim:2018eyx}.
Figure~\ref{fig:Ti-beta} shows with thick black lines the parameter space reproducing the observed DM density for different DM masses.
The shaded regions represent areas constrained by different observables: $\Min\lesssim 10^{-1}$~g and $\Min\gtrsim 2\times 10^8$~g are disfavored by CMB and BBN (both in red), large values for $\beta$ (in green) lead to GW backreaction (Eq.~\eqref{eq:GW}) or to a universe dominated by GWs at the onset of the BBN era (Eq.~\eqref{eq:GW2}).
Additionally, $\beta\gtrsim 10^{-5}$ tend to produce hot DM, however, it is important to note that the latter constraint {\it only} applies to the case of light DM ($\mdm \ll \Tbh$), with mass $\mdm \lesssim 30$~GeV.
Finally, $\beta$ values smaller than $\sim 10^{-24}$ can not accommodate the total observed DM abundance.

It is important to recall that, in this figure, the effects of both the SM-DM interactions and the DM self-interactions have been neglected, and therefore it corresponds to DM produced solely via Hawking evaporation of PBHs.
The thick black lines show three different slopes, corresponding to three different regimes.
If PBHs dominated the universe energy density, the DM yield is independent of $\beta$, and therefore the lines are vertical.
In this regime, $\mdm\simeq 10^9$~GeV is the lowest viable DM mass.
However, a $\beta$ dependence shows up if, during the whole BH lifetime, the universe was radiation dominated.
In this case, two regimes arise, depending on whether DM is lighter or heavier than the initial BH temperature, Eq.~\eqref{eq:N}.
In the former case $\beta\propto\Tin$, whereas in the latter $\beta\propto\Tin^{-3}$.
Finally, let us note that in the present case where DM self-interactions are not efficient, DM has to be heavier than $\mathcal{O}(1)$~MeV in order not to be hot, and can be as heavy as $M_P$~\cite{Chung:1999ve, Giudice:1999fb} (notice that we are not considering a BH evaporation process stopping at $\Tbh\sim M_P$, with the associated production of Planck mass relics~\cite{MacGibbon:1987my, Barrow:1992hq, Carr:1994ar, Dolgov:2000ht, Baumann:2007yr, Hooper:2019gtx}).

%%%%%%%%%%%%%%%%%%%%%%%%%%%%%%%%%%%%%%%%%%%%%%%%%%%%%%%%%%
\section{Self-interacting Dark Matter from Primordial Black Holes} \label{sec:SI}
%%%%%%%%%%%%%%%%%%%%%%%%%%%%%%%%%%%%%%%%%%%%%%%%%%%%%%%%%%
In the previous section, the production of {\it collisionless} DM particles via the evaporation of PBHs was presented.
However, DM can feature sizable {\it self-interactions}, dramatically changing its expected relic density.
To analytically understand the role played by DM self-interactions~\cite{Bernal:2020gzm}, let us study the DM production under the assumption of instantaneous evaporation.

%%%%%%%%%%%%%%%%%%%%%%%%%%%%%%%%%%%%%%%%%%%%
\subsection{Radiation Dominated Universe}
First, we focus on the case where the universe was dominated by SM radiation during the whole lifetime of PBHs.
In the case where the DM is lighter than the initial BH temperature, each BH radiates $\Ndm$ DM particles with a mean energy $\langle E\rangle \simeq 6\,\Tbh$.
The total DM energy density radiated by a BH can be estimated by
\begin{equation}
    \rho_\text{DM}(T=\Tev) \simeq \ndm(T=\Tev)\,\Tbh
    = \beta\,\frac{\gdm\,\zeta(3)\,\Cn}{2\pi^2}\,\Tin\,\Tev^3\,,
\end{equation}
where the SM entropy conservation and Eq.~\eqref{eq:N} were used.
Let us notice that the produced DM population inherits the BH temperature, and, as DM is {\it not} in chemical equilibrium, develops a large chemical potential.

If the dark sector has sizable self-interactions guaranteeing that elastic scatterings within the dark sector reach kinetic equilibrium, DM thermalizes with a temperature $T'$ in general different from the SM one $T$ and the BH temperature $\Tbh$.
Thermalization within the dark sector is guaranteed if DM reaches kinetic equilibrium, i.e., if the rate of DM elastic scattering is higher than the Hubble expansion rate.
Additionally, if number-changing self-interactions within the dark sector reach chemical equilibrium, the DM number density just after thermalization is therefore
\begin{equation}\label{eq:ndmboostlightRD}
    \ndm(\Tpev) = 
    \begin{cases}
        \frac{15^{3/4}\zeta(3)^{7/4}\,\gdm\,\Cn^{7/4}}{\pi^5\,\Cr^{3/4}} \beta^{3/4} \left(\Tin\,\Tev^3\right)^{3/4}\qquad&\text{for}\quad \mdm\ll\Tpev\,,\\[10pt]
        \beta\,\frac{\gdm\,\zeta(3)\,\Cn}{2\pi^2}\,\frac{\Tin\,\Tev^3}{\mdm}\qquad&\text{for}\quad \mdm\gg\Tpev\,,
    \end{cases}
\end{equation}
where $\Cr = 1$ (bosonic DM) or $7/8$ (fermionic DM).
The overall effect of self-interactions and in particular of number-changing interactions within the dark sector is to decrease the DM temperature, increasing the DM number density.

%%%%%%%%%%%%%%%%%%%%%%%%%%%%%
\subsection{Matter Dominated Universe}
%%%%%%%%%%%%%%%%%%%%%%%%%%%%%
Now, we focus on the alternative scenario, in which PBHs at some point dominated the total energy density.
In this case, the total DM energy density for light DM particles radiated by a PBH is
\begin{equation}
    \rho_\text{DM}(T=\Tbev) \simeq \frac{\gdm\,\zeta(3)\,\Cn}{2\pi^2}\,\Tbev^4\,,
\end{equation}
and therefore, the DM number density just after thermalization becomes
\begin{equation}
    \ndm(\Tpev) \simeq
    \begin{cases}
        \frac{\zeta(3)\,\Cn}{\pi^2}\gdm\left(\frac{15\,\zeta(3)\,\Cn}{\pi^4\,\Cr}\right)^{3/4} \Tbev^3\qquad&\text{for}\quad \mdm\ll\Tpev\,,\\[8pt]
        \frac{\gdm\,\zeta(3)\,\Cn}{2\pi^2}\,\frac{\Tbev^4}{\mdm}\qquad&\text{for}\quad \mdm\gg\Tpev\,.
    \end{cases}
\end{equation}

Finally, we note that for DM heavier than the initial PBH mass, thermalization and number-changing interactions do not play a role, since the originally Hawking radiated particles were almost non-relativistic

%%%%%%%%%%%%%%%%%%%%%%%%%%%%%%%%%%%%%%%%%%%%%%%%%%%%%%
\begin{figure}
	\centering
	\includegraphics[scale=0.58]{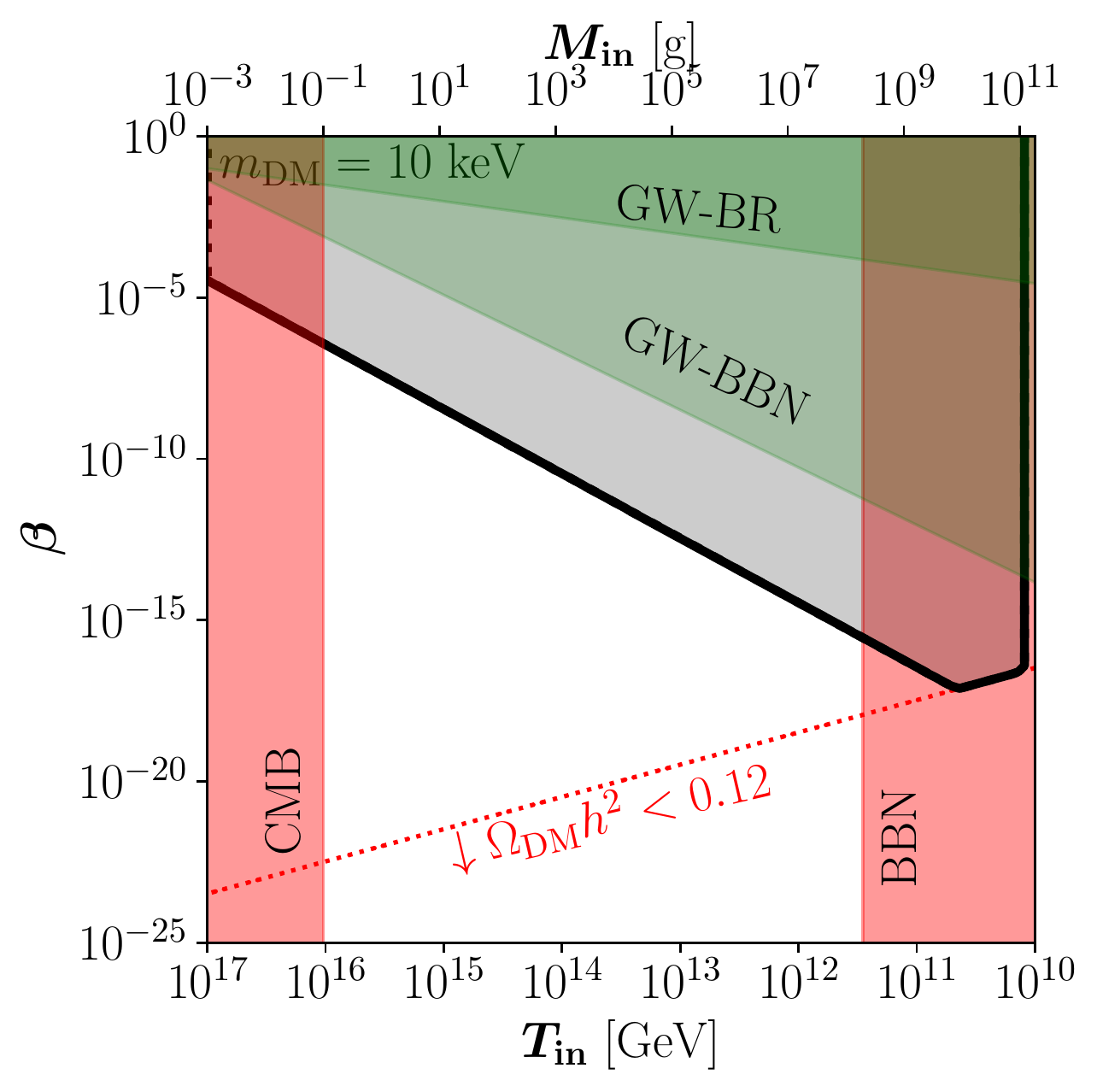}
	\includegraphics[scale=0.58]{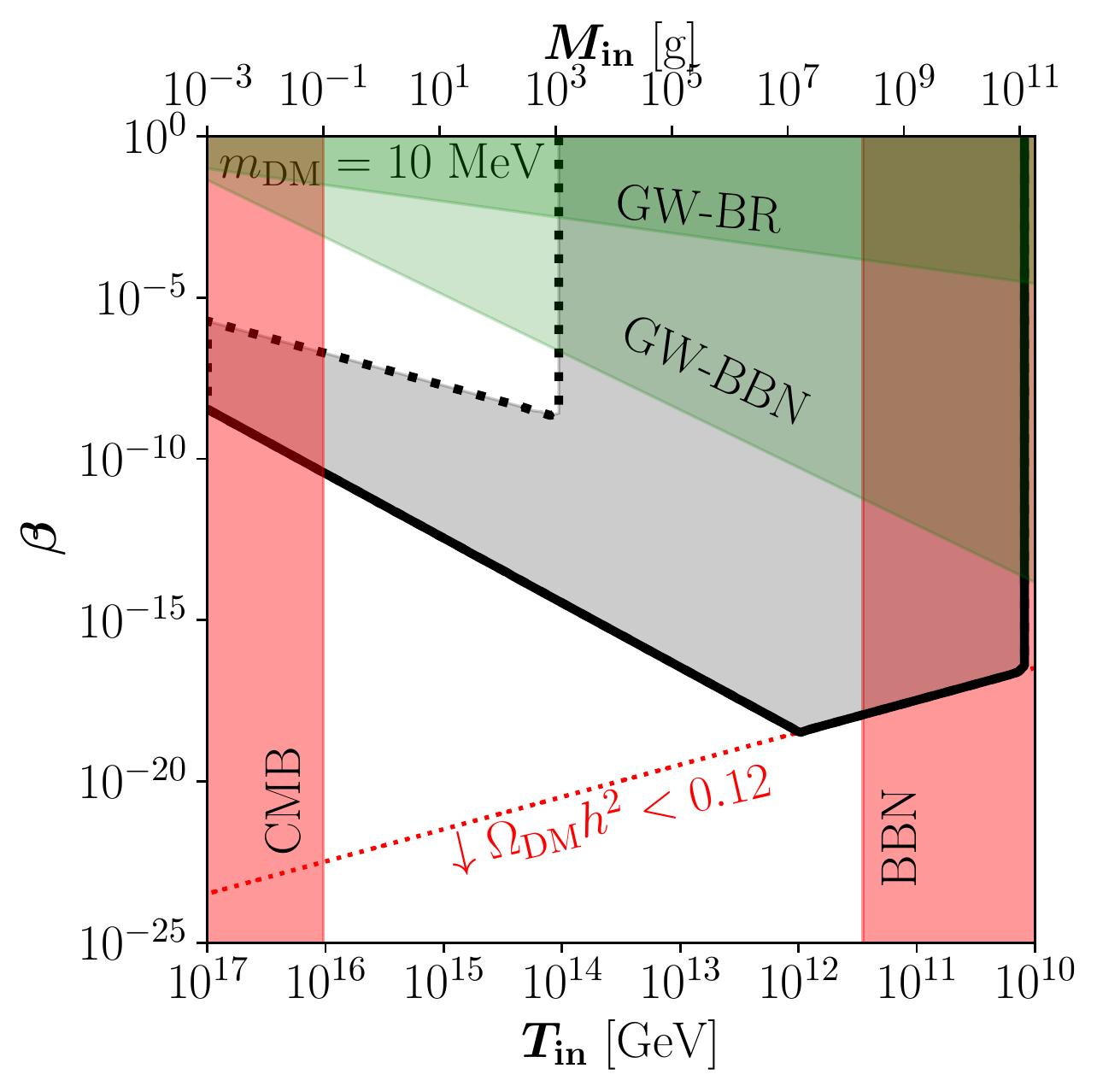}
	\includegraphics[scale=0.58]{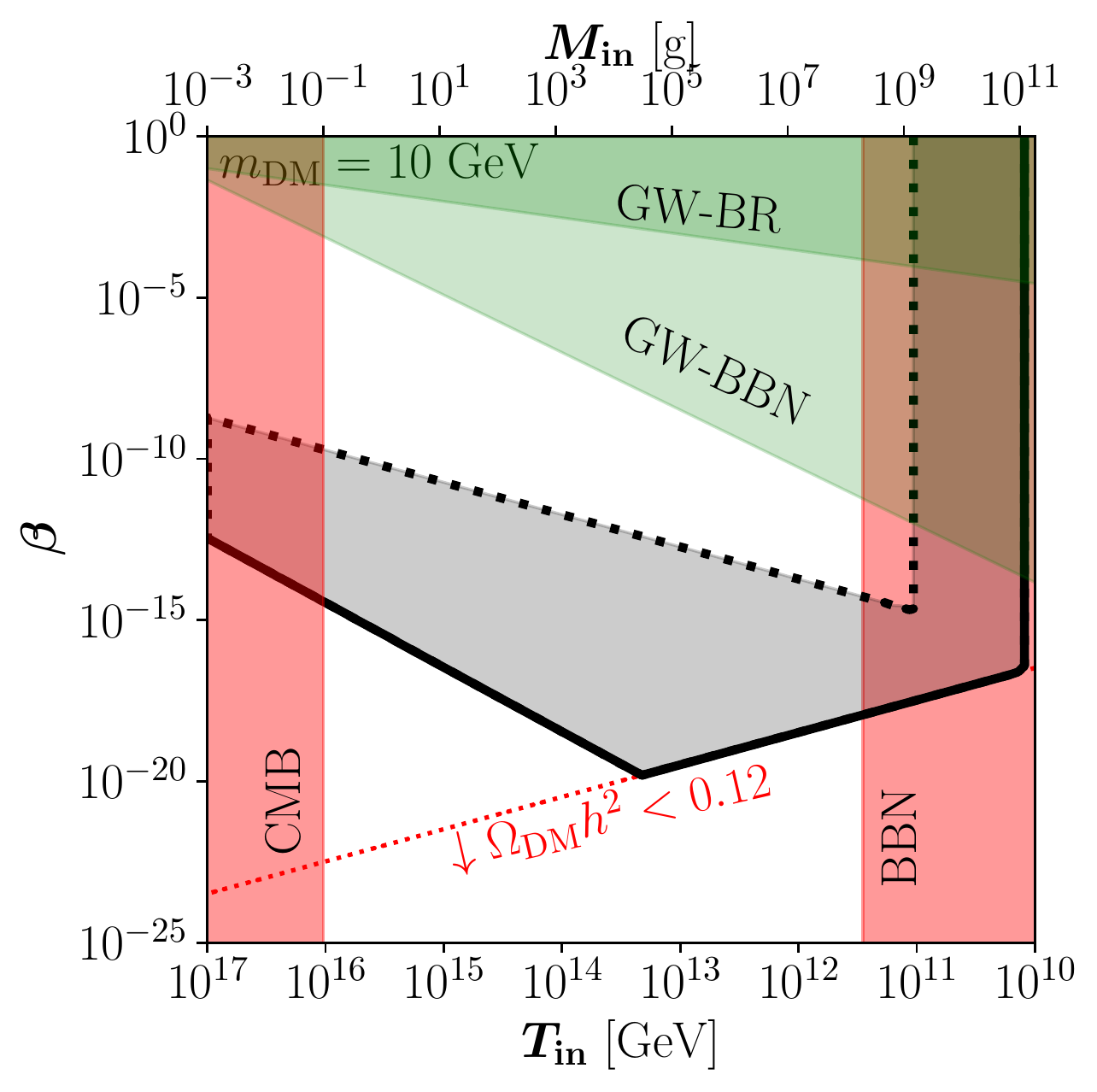}
	\includegraphics[scale=0.58]{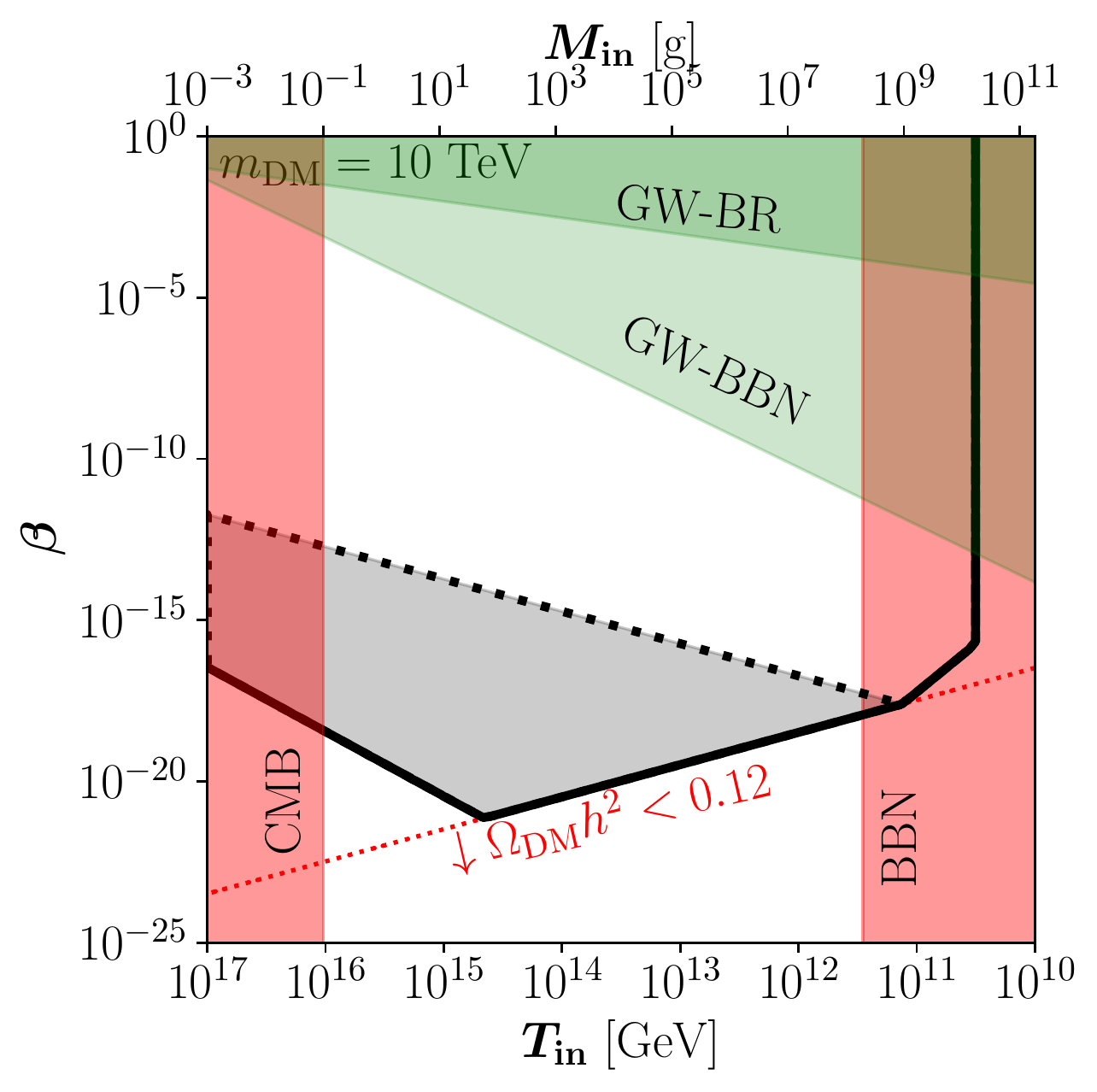}
	\caption{Parameter space reproducing the observed DM abundance (shaded gray areas) from PBH evaporation, for different DM masses.
	The thick black lines show the limiting cases without (dotted lines) and with a maximum effect from DM self-interactions (solid lines).
	The shaded green and red areas are excluded by different observables and described in the text.}
	\label{fig:Ti-beta-masses}
\end{figure} 
%%%%%%%%%%%%%%%%%%%%%%%%%%%%%%%%%%%%%%%%%%%%%%%%%%%%%%
The impact of DM self-interactions is shown in Fig.~\ref{fig:Ti-beta-masses}, for different DM masses.
Dotted and solid thick lines correspond to the limiting cases without and with a maximal effect from self-interactions, respectively, in the same parameter space used in Fig.~\ref{fig:Ti-beta}.
Out of the six regimes presented previously, four are visible in the plots and are described in ascending order for $\Tin$.
\begin{itemize}
    \item The observed DM abundance can be generated in the case where the PBHs dominate the universe energy density (above the red dotted line), only for heavy DM, i.e. $\mdm > \Tbhin$.
    The DM yield is independent from $\beta$ and there is no boost due to self-interactions, as can be seen in the lower right panel corresponding to $\mdm=10$~TeV.
    We notice that this scenario is typically excluded by the BBN constraint, and only viable for $\mdm\gtrsim 10^9$~GeV. 
    \item The case of heavy DM, this time in a radiation dominated scenario, appears when $\mdm\gtrsim 100$~GeV.
    This case is visible in the lower right panel, where the DM yield is given by Eq.~\eqref{eq:YdmRD} without a significant boost. It follows that $\beta\propto\Tin^{-3}$.
    This scenario is again typically in tension with the BBN observations, and only viable when $\mdm\gtrsim 10^5$~GeV.
    \item The third regime corresponds to $\Tpev\ll\mdm\ll\Tbhin$, where the DM yield in Eq.~\eqref{eq:YdmRD} is boosted by a factor $\Tbhin/\mdm$.
    In this case, $\beta \propto \Tin^{-1}$.
    We notice that values of $\beta$ smaller than the ones required in this scenario always produce a DM underabundance.
    \item The last regime happens for light DM $\mdm\ll\Tpev$ in a radiation-dominated universe.
    The DM yield in Eq.~\eqref{eq:YdmRD} is boosted by a factor $\propto \Tbhin/(\Tin^{1/4}\,\Tev^{3/4})$.
    In this case, the DM abundance requires $\beta \propto \Tin^2$.
\end{itemize}

%%%%%%%%%%%%%%%%%%%%%%%%%%%%%%%%%%%%%%%%%%%%%%%%%%%%%%%%%%
\section{Other Purely Gravitational Dark Matter Production Modes} \label{sec:gravity}
%%%%%%%%%%%%%%%%%%%%%%%%%%%%%%%%%%%%%%%%%%%%%%%%%%%%%%%%%%
Until now, we have focused on DM particles produced in the early universe by the gravitational Hawking evaporation of PBHs.
This process can be the dominant in the case where other portals are closed.
However, even in this case, there are other purely gravitational channels that can not be neglected.
In the following we describe the 2-to-2 DM production via the scattering of SM particles or inflatons, via the $s$-channel exchange of gravitons.

%%%%%%%%%%%%%%%%%%%%%%%%%%%%%%%%%%%%%%%%%%%%%%%%%%%%%%%%%%
\subsection{Dark Matter from Standard Model Scatterings}
%%%%%%%%%%%%%%%%%%%%%%%%%%%%%%%%%%%%%%%%%%%%%%%%%%%%%%%%%%
Independently from the PBH evaporation, there is an irreducible DM production channel which is particularly efficient in the region favored by Fig.~\ref{fig:Ti-beta}, and corresponds to the gravitational UV freeze-in.
DM can be generated via 2-to-2 annihilations of SM particles, mediated by the exchange of massless gravitons in the $s$-channel~\cite{Garny:2015sjg, Tang:2017hvq, Garny:2017kha, Bernal:2018qlk}.
In the case $\mdm \ll \Trh$,  its contribution to the total DM density is
\begin{equation} \label{eq:FIlight}
    Y_\text{DM} = \frac{45\, \alpha_\text{DM}}{2\pi^3\, \gss} \sqrt{\frac{10}{\gs}} \left(\frac{\Trh}{M_P}\right)^3,
\end{equation}
where $\alpha_\text{DM}=1.9\times 10^{-4}$, $1.1\times 10^{-3}$ or $2.3\times 10^{-3}$ for scalar, fermionic, or vector DM, respectively.
Instead, if the produced particle is heavier than the reheating temperature (but still below $\Tmax$), it cannot be generated after but during reheating. In that case, the yield can be computed in the range $\Tmax \geq T \geq \mdm$:
\begin{equation} \label{eq:FIheavy}
    Y_\text{DM} = \frac{45\, \alpha_\text{DM}}{2\pi^3\, \gss} \sqrt{\frac{10}{\gs}} \frac{\Trh^7}{M_P^3\, \mdm^4}\,.
\end{equation}
Away from the instantaneous decay approximation, the DM yield is only boosted by a small factor of order $\mathcal{O}(1)$ for an inflaton behaving as non-relativistic matter~\cite{Garcia:2017tuj, Bernal:2019mhf}. 
Nevertheless, it is important to note that a big boost factor can appear when considering nonthermal effects~\cite{Garcia:2018wtq}, or expansion eras dominated by a fluid component stiffer than radiation~\cite{Bernal:2019mhf, Bernal:2020bfj}.

\begin{figure}
	\centering
	\includegraphics[scale=0.58]{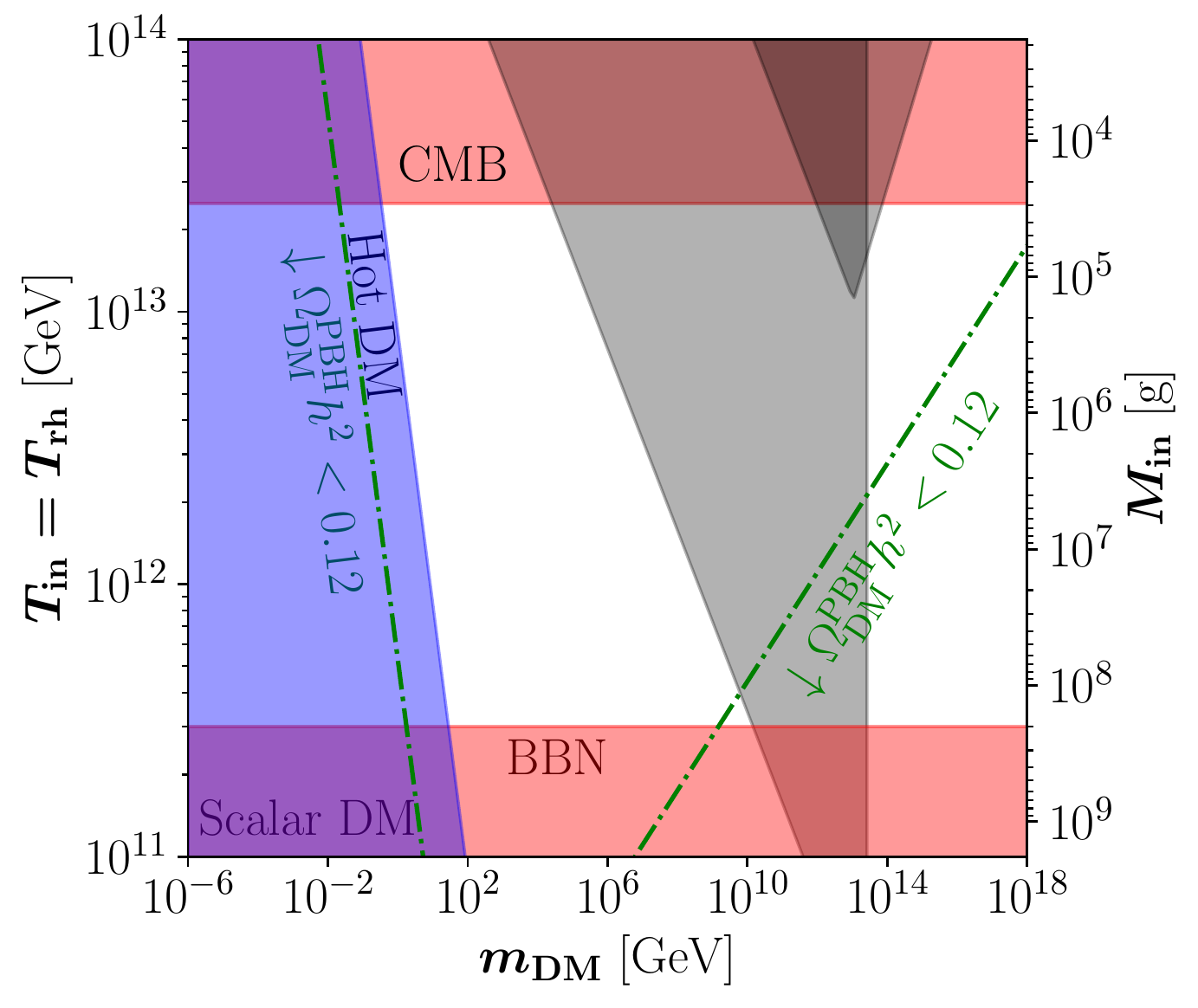}
	\includegraphics[scale=0.58]{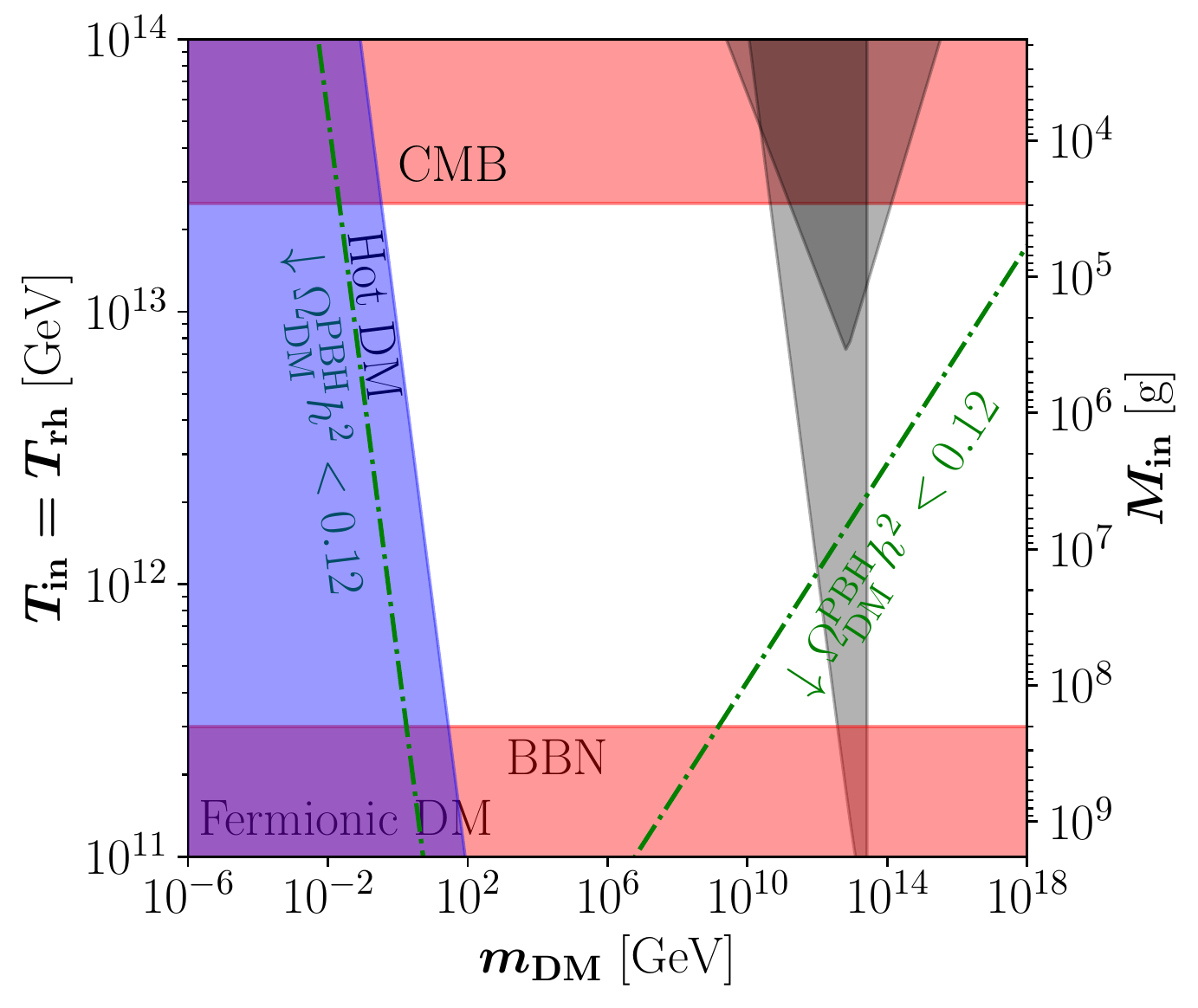}
	\includegraphics[scale=0.58]{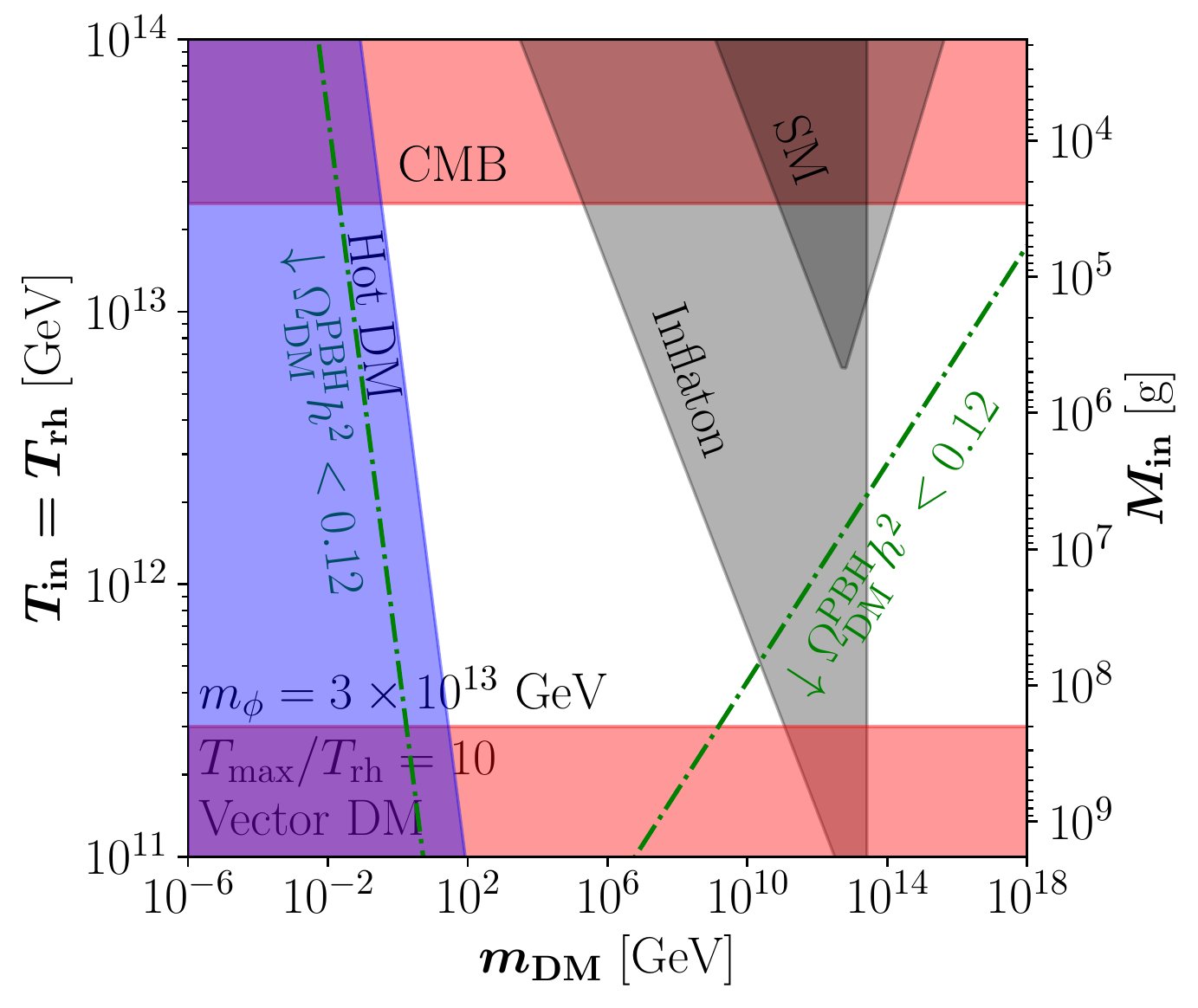}
	\caption{Parameter space reproducing the observed DM abundance (shaded gray areas) from PBH evaporation, for different DM masses.
	The thick black lines show the limiting cases without (dotted lines) and with a maximum effect from DM self-interactions (solid lines).
	The shaded green and red areas are excluded by different observables and described in the text.}
	\label{fig:Ti-beta-grav}
\end{figure}

The {\it borders} of the gray regions labeled `SM' in Fig.~\ref{fig:Ti-beta-grav} show the parameter space compatible with the measurements of DM relic abundance, for DM produced via SM scatterings mediated by gravitons.
Additionally, the gray {\it regions} generate a DM overabundance, being therefore excluded.
We note that the temperatures $\Trh$ and $\Tin$ are in principle unrelated, however, $\Trh \geq \Tin$ to guarantee that PBHs are produced after the onset of the radiation domination era.
Assuming $\Trh = \Tin$, Fig.~\ref{fig:Ti-beta-grav} also shows the constraint on PBHs coming from CMB, BBN (both in red) and hot DM (blue).
Below the dash-dotted lines, PBHs cannot generate the whole observed DM abundance.
On the contrary, in the white area (and above the dash-dotted lines), the whole DM relic density can be radiated by PBHs, for a given value of $\beta$.
%%%%%%%%%%%%%%%%%%%%%%%%%%%%%%%%%%%%%%%%%%%%%%%%%%%%%%
%%%%%%%%%%%%%%%%%%%%%%%%%%%%%%%%%%%%%%%%%%%%%%%%%%%%%%

%%%%%%%%%%%%%%%%%%%%%%%%%%%%%%%%%%%%%%%%%%%%%%%%%%%%%%%%%%
\subsection{Dark Matter from Inflaton Scatterings}
%%%%%%%%%%%%%%%%%%%%%%%%%%%%%%%%%%%%%%%%%%%%%%%%%%%%%%%%%%
Alternatively to the production via SM scatterings, DM can also be generated by 2-to-2 inflaton scatterings mediated by the $s$-channel exchange of a graviton during the reheating era.
The yield $Y_\text{DM}$ at the end of reheating (i.e., at $T = \Trh$) can be analytically computed and reads~\cite{Mambrini:2021zpp, Bernal:2021kaj, Barman:2021ugy}
\begin{equation} \label{eq:Y0}
    Y_\text{DM} = \frac{\gs^2}{81920\, \gss} \sqrt{\frac{10}{\gs}}\, \left(\frac{\Trh}{M_P}\right)^3 \left[\left(\frac{\Tmax}{\Trh}\right)^4 - 1\right] f\left(\frac{\mdm}{m_\phi}\right).
\end{equation}
where $m_\phi$ is the inflaton mass, and we have defined
\begin{equation}
f(x) \equiv
    \begin{cases}
        \left(x^{2}+2\right)^{2}\sqrt{1-x^{2}} & \mbox{for real scalars},\\[8pt]
        x^{2}\left(1-x^{2}\right)^{3/2} & \mbox{for Dirac fermions},\\[8pt]
        \frac{1}{8}\sqrt{1-x^2}\left(4+4x^2+19x^4\right)& \mbox{for vector bosons.}
    \end{cases}
\end{equation}

The {\it borders} of the gray regions labeled `Inflaton' in Fig.~\ref{fig:Ti-beta-grav} show again the parameter space compatible with the measurements of DM relic abundance, for DM produced via inflaton scatterings mediated by gravitons, assuming $\Tmax/\Trh = 10$ and $m_\phi = 3 \times 10^{13}$~GeV.
The strong cuts at $\mdm = m_\phi$ correspond to the kinematical production threshold.
We notice that this two purely gravitational production modes set strong constraints, excluding large regions of the parameter space favored by PBHs.
In particular, PBHs evaporating before the onset of BBN cannot be the responsible for the genesis of DM with masses on the range $\sim 10^{10}$~GeV to $\sim 10^{13}$~GeV for scalar and vector DM, or the range $\sim 10^{10}$~GeV to $\sim 10^{13}$~GeV for fermionic DM.

%%%%%%%%%%%%%%%%%%%%%%%%%%%%%%%%%%%%%%%%%%%%%%%%%%%%%%%%%%
\section{Conclusions} \label{sec:conclusions}
%%%%%%%%%%%%%%%%%%%%%%%%%%%%%%%%%%%%%%%%%%%%%%%%%%%%%%%%%%
Even if it is usually assumed that the visible and dark sectors are connected via interactions at the electroweak scale (as in the case of WIMPs), this should not be the case.
Interactions could be much smaller, so that DM never reaches thermal equilibrium with the standard model, FIMPs being an example.
However, what is definitely guaranteed is that gravity is the {\it minimal} interaction mediating between the two sectors.
Therefore, gravity will {\it unavoidable} produce DM in the early universe, and could even be responsible for the whole observed DM abundance.

One particularly interesting channel for DM produced gravitationally corresponds to the Hawking evaporation of PBHs.
During their evaporation, they radiate all particles including states of the dark sector, potentially generating all relic DM abundance.
However, if the dark sector features sizable (i.e., non-gravitational) self-interactions, thermalization and number-changing processes can have a strong impact, in particular enhancing the produced DM relic abundance by several orders of magnitude.
Here we have estimated the boost from general arguments such as the conservation of energy and entropy, independently from the underlying particle physics details of the dark sector.
Two main consequences can be highlighted: $i)$ As the DM abundance is increased, a smaller initial energy density of PBHs is required.
$ii)$ Thermalization in the dark
sector decreases the mean DM kinetic energy, relaxing the bound from structure formation and hence, allowing for lighter DM in the keV ballpark.

However, there are other {\it irreducible} gravitational channels.
We studied the DM production by 2-to-2 scattering of SM particles or inflatons, mediated by the $s$-channel exchange of a graviton.
In particular, we have shown that all previously mentioned gravitational production channels are effective in the same region of the parameter space, i.e., for heavy DM and high reheating temperatures.
Therefore, the interplay of the different mechanisms set strong bonds, and exclude large regions of the parameter space favored by the PBH production.

%%%%%%%%%%%%%%%%%%%%%%%%%%%%%%%%%%%%%%%%%%%%%%%%%%%%%%%%%%
\section*{Acknowledgments}
%%%%%%%%%%%%%%%%%%%%%%%%%%%%%%%%%%%%%%%%%%%%%%%%%%%%%%%%%%
This article is based on the talk given in the BSM-2021 at Zewail City.
The original works~\cite{Bernal:2020kse, Bernal:2020ili, Bernal:2020bjf} were done in collaboration with Óscar Zapata.
The author received funding from Universidad Antonio Nariño grants 2019101 and 2019248, the Spanish FEDER/MCIU-AEI under grant FPA2017-84543-P, and the Patrimonio Autónomo - Fondo Nacional de Financiamiento para la Ciencia, la Tecnología y la Innovación Francisco José de Caldas (MinCiencias - Colombia) grant 80740-465-2020.
This project has received funding/support from the European Union's Horizon 2020 research and innovation programme under the Marie Skłodowska-Curie grant agreement No 860881-HIDDeN.

\bibliographystyle{unsrt}
\bibliography{biblio}
\end{document}